\documentclass[
    ,final            
  ]
  {aipproc}

\layoutstyle{6x9}

\begin{document}

\title{Non-Gaussian Effects in Identical Pion Correlation Function at STAR}


\author{M.~Bysterský (for the STAR Collaboration)}{
  address={
Nuclear Physics Institute, 
Academy of Sciences of the Czech Republic,
250 68 \v Re\v z near Prague, Czech Republic}
}

\begin{abstract}
Preliminary femtoscopic results on identical pions
from high statistics data set of Au+Au collisions at 
$\sqrt{s_{{\rm NN}}}=200~{\rm GeV}$ taken
during the fourth RHIC run are presented.
The measured three-dimensional correlation function
is studied at low relative momenta using the 
Gaussian parametrization and the Lévy stable parametrization.
The latter is expected to better describe the data.
As the results show, both parametrizations 
underestimate the peak of the measured
correlation function equally. 
\end{abstract}

\maketitle


\section{Introduction}

Motivation for this study is to check whether a 
recently proposed parametrization of the correlation function
using Lévy stable source distribution~\cite{Csorgo:2003uv} 
brings a significant improvement over 
the standard Gaussian fit~\cite{Wiedemann:1999qn}. 
In addition to this, the non-Gaussian source distribution function 
is also used in most of the models, but the standard 
method of fitting experimental correlation function assumes a 
Gaussian source~\cite{Wiedemann:1999qn}.

Possible methods of studying the non-Gaussian effects in 
the experimental correlation function include 
Edgeworth expansion~\cite{Csorgo:2000pf} and 
Lévy stable source distribution parametrization~\cite{Csorgo:2003uv}.

\section{Correlation function of identical pions}

\paragraph{Event and particle selection criteria}

We briefly list the values of the event and particle 
selection criteria used in the present analysis. Detailed 
description can be found in~\cite{Adams:2004yc}. 
Events are 
binned by centrality in 
five bins corresponding to 
0--5\%, 5--10\%, 10--20\%, 20--30\% and 30--80\%
of the total hadronic Au+Au cross-section. 
In addition to the track cuts listed in~\cite{Adams:2004yc}, 
measured specific ionization of pions 
is required to be farther than $\pm 2 \sigma$ 
from the Bethe-Bloch theoretical value for electrons.
This cut removes a contamination due to 
conversion electrons in the low momentum region.
Pairs of identical pions are binned by average transverse
pair momentum $k_{\rm T} = \frac{1}{2} |\vec{p}_{\rm T1}+\vec{p}_{\rm T2}|$
in four bins corresponding to 
$k_{\rm T} \in(0.15$--$0.25, 0.25$--$0.35, 0.35$--$0.45,
0.45$--$0.60)~{\rm GeV}/c$ and the results are presented 
as a function of the average $k_{\rm T}$ in each of these bins.
\\

Experimentally, two-particle correlations are studied 
by constructing the correlation function as a ratio
\begin{equation}
  \label{eq:Cdefinition}
  C(\vec{q}) = \frac{A(\vec{q})}{B(\vec{q})}\,,
\end{equation} 
where $A(\vec{q})$ is the measured distribution of the momentum
difference $\vec{q} = \vec{p}_{1}-\vec{p}_{2}$ for pairs of particles
from the same event and $B(\vec{q})$ is the corresponding 
reference distribution for pairs of particles 
from different events belonging to the same event class as 
analyzed event~\cite{Wiedemann:1999qn}.
The two-particle correlation function 
at low relative momentum $\vec{q}$ of identical pion pairs 
is studied using Bertsch-Pratt 
parametrization~\cite{Podgoretsky:1982xu,Pratt:1986ev,Bertsch:1989vn}
in the longitudinal co-moving system (LCMS) frame, where
the relative momentum vector is decomposed into 
the out, side and long components
$\vec{q}=(q_{\rm o}, q_{\rm s}, q_{\rm l})$.

\subsection{Gaussian parametrization}

Standard method of fitting the two-pion 
correlation function assumes the Gaussian 
source distribution.
The correlation function is usually parametrized
by a three-dimensional Gaussian in 
$\vec{q}$~\cite{Wiedemann:1999qn}.
Taking into account a repulsive Coulomb
interaction between charged identical pions, 
the measured correlation function~\eqref{eq:Cdefinition} 
is fitted using Bowler-Sinyukov 
procedure~\cite{Bowler:1991vx,Sinyukov:1998fc},
\begin{equation}
  \label{eq:Gaussian}
  C(\vec{q}) = 
  (1 - \lambda)
  + \lambda K_{\rm c}
  \bigg[
  1 + \exp 
  \Big(-\sum_{\rm i,j}R_{\rm ij}^{2}\,q_{\rm i}q_{\rm j}
  \Big)
  \bigg]\,.
\end{equation} 
Here, the correlation strength $\lambda$ equals the
fraction of pairs originating in the same spatio-temporal 
region relevant for Bose-Einstein correlations, 
$K_{\rm c}$ is the squared Coulomb wave-function integrated
over the source with radius $5~{\rm fm}$~\cite{Adams:2004yc}.
$R_{\rm ij}$ are the Gaussian source radius 
parameters defined as the widths of the source emission function.
Let us note, that
only the pairs obeying Bose-Einstein statistics 
are considered to Coulomb interact.
For an azimuthally integrated analysis the cross-terms 
vanish~\cite{Wiedemann:1999qn}, 
$R_{\rm ij}=0,\ {\rm i} \neq {\rm j}$.

\subsection{Non-Gaussian parametrization}

Detailed analysis of the shape of the correlation function 
is important because it carries information about the 
space-time structure of the particle 
emitting source~\cite{Csorgo:2003uv,Wiedemann:1999qn}.
Deviations from Gaussian shape can be studied using
Edgeworth expansion or Lévy stable source distribution.

Edgeworth expansion~\cite{Csorgo:2000pf} provides
model-independent approach for an analysis of 
the shape of the correlation function. 
In our previous pion interferometry 
analysis~\cite{Adams:2004yc}
it was shown that the Edgeworth expansion,
based on an experimentally preferred Gaussian
weight function and a complete orthogonal set of 
even order Hermite polynomials, 
up to $6^{\rm th}$ order is sufficient to describe the data.
However, physical interpretation of the higher order 
($4^{\rm th}$, $6^{\rm th}$) fit parameters is not clear.

\subsubsection{Lévy stable source parametrization}

To study possible deviations of the correlation function
from a Gaussian shape, 
we followed the method suggested in~\cite{Csorgo:2003uv}.
This formalism is relevant for the femtoscopic 
studies of the expanding systems created 
in heavy-ion collisions, where the scale of the 
fluctuations may be characterized by long tails 
and asymptotic power-law like behavior.
The probability distribution of particle emission
points then corresponds to a 
Lévy stable distribution~\cite{Csorgo:2003uv}.
Using the Bowler-Sinyukov procedure
to determine the repulsive Coulomb
interaction between identical pions,
the two-particle correlation function
is then characterized by a stretched 
exponential parametrization, 
\begin{equation}
  \label{eq:Levy}
  C(\vec{q}) = 
  (1 - \lambda)
  + \lambda K_{\rm c}
  \Bigg[
  1 + \exp 
  \bigg(-
  \Big(
  \sum_{\rm i,j}R_{\rm ij}^{2}\,q_{\rm i}q_{\rm j}
  \Big)^{\alpha / 2}
  \bigg)
  \Bigg]\,.
\end{equation} 
The additional parameter $\alpha$,
when compared to~\eqref{eq:Gaussian}, 
is the Lévy index of stability, $0 < \alpha \le 2$.
For $\alpha = 2$ the Gaussian form~\eqref{eq:Gaussian} 
is recovered, while for $\alpha < 2$ the correlation 
function becomes more peaked than a Gaussian and
develops longer tails.
For the azimuthally integrated analysis, 
$R_{\rm ij}=0,\ {\rm i} \neq {\rm j}$.

\section{Discussion of Results}

Here we present results on two-particle 
correlations of charged identical pions 
in Au+Au collisions at $\sqrt{s_{{\rm NN}}}=200~{\rm GeV}$ measured 
in the STAR detector during the fourth RHIC run (2003--2004).
Data set of $11 \times 10^6$ minimum-bias triggered 
events is used for this analysis.

The interferometric parameters, 
correlation strength $\lambda$ and 
radii $R_{\rm o}$, $R_{\rm s}$ and $R_{\rm l}$, are obtained
by fitting the measured correlation function~\eqref{eq:Cdefinition}
with the Gaussian parametrization~\eqref{eq:Gaussian}. 
Interferometric radius parameters measure the sizes
of the homogeneity regions,
regions from where the particles are emitted with 
the same average pair momentum $k_{\rm T}$, and their 
$k_{\rm T}$ dependence contains dynamical information of
the pion emitting source~\cite{Wiedemann:1999qn}.

\begin{figure}
  \includegraphics[height=.5\textheight]{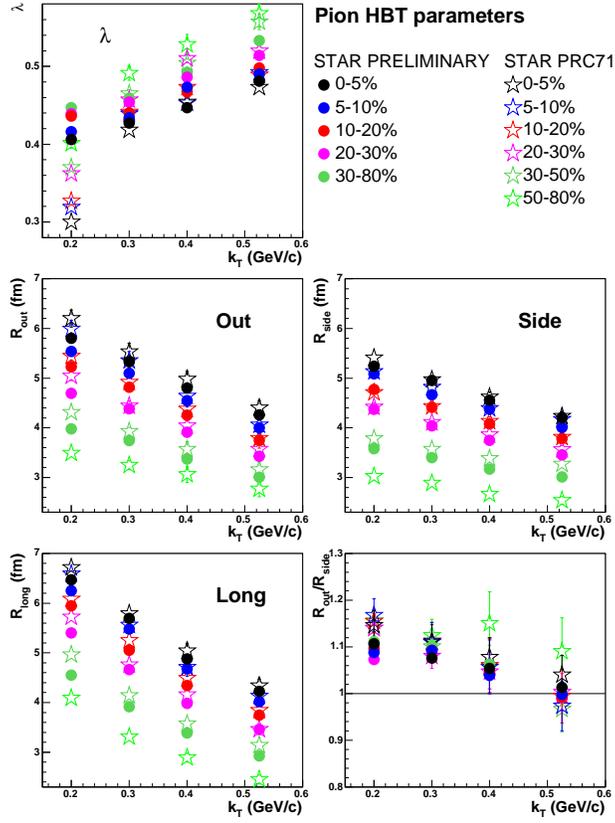}
  \caption{
    Interferometric parameters as functions of 
    $k_{\rm T}$ and centrality. Results of 
    the Gaussian parametrized fit~\eqref{eq:Gaussian}
    to the experimental correlation function, 
    Au+Au collisions at $\sqrt{s_{{\rm NN}}}=200~{\rm GeV}$.
  }
  \label{STAR_PRC71}
\end{figure}

Figure~\ref{STAR_PRC71} shows STAR preliminary 
results on interferometric parameters 
as functions of $k_{\rm T}$ for five centrality bins,
where the three-dimensional experimental correlation function 
is fitted using the Gaussian parametrization of the 
correlation function~\eqref{eq:Gaussian}.
Results of the analysis of higher statistics data set of
Au+Au collisions at $\sqrt{s_{{\rm NN}}}=200~{\rm GeV}$
are compared with
the previously analyzed STAR data~\cite{Adams:2004yc}
for the same system at the same energy 
taken during the second RHIC run (2001--2002).
It can be seen that the extracted interferometric radii
$R_{\rm o}$, $R_{\rm s}$ and $R_{\rm l}$ are consistent
within errors with the previous analysis. 
A small systematic shift in radii
can be attributed to the momentum resolution
correction which is not included in this analysis. 
Significant difference in $\lambda$ 
is observed in the lowest $k_{\rm T}$ bin.
This is explained by an improved purity of the pion sample,
where the additional cut on particle specific ionization 
removes contamination from the conversion electrons 
in the low momentum region.

\begin{figure}
  \includegraphics[height=.35\textheight]{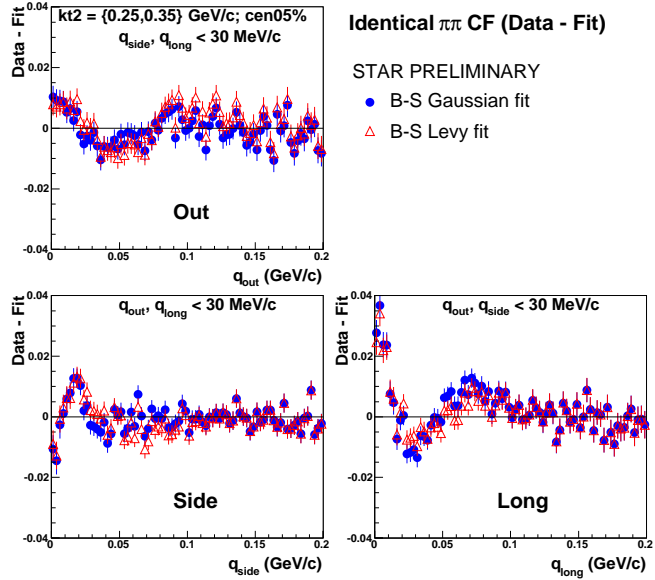}
  \caption{
    Gaussian parametrized fit~\eqref{eq:Gaussian} to data 
    compared to the Lévy stable source parametrized fit~\eqref{eq:Levy} 
    to data, each subtracted from the 
    experimental correlation function.
  }
  \label{BS-Levy-DataMinusFit}
\end{figure}

The Gaussian parametrized fit~\eqref{eq:Gaussian}
and the Lévy stable 
source parametrized fit~\eqref{eq:Levy} 
to the measured correlation function,
each subtracted from the measured 
three-dimensional correlation function,
are projected in the out, side and long  coordinates 
and compared in Figure~\ref{BS-Levy-DataMinusFit}.
Correlation function is shown for the
$k_{\rm T}$ bin $0.25$--$0.35~{\rm GeV}/c$,
for the most central collisions
and the projections are constrained 
by the unprojected variables 
$q_{\rm o}, q_{\rm s}, q_{\rm l} < 30~{\rm MeV}/c$.
It can be seen that contrary to~\cite{Csorgo:2003uv},
Lévy stable source 
distribution parametrization~\eqref{eq:Levy}
does not fit the three-dimensional 
experimental correlation function
significantly better
than the standard Gaussian 
parametrization~\eqref{eq:Gaussian}.
Both parametrizations
equally underestimate the peak value of the measured
correlation function for relative momenta
$q_{\rm o}, q_{\rm s}, q_{\rm l} < 20~{\rm MeV}/c$
and underestimate the tail of the correlation function.
Both effects are mostly visible in the 
long projection in Figure~\ref{BS-Levy-DataMinusFit}.

\begin{figure}
  \includegraphics[height=.5\textheight]{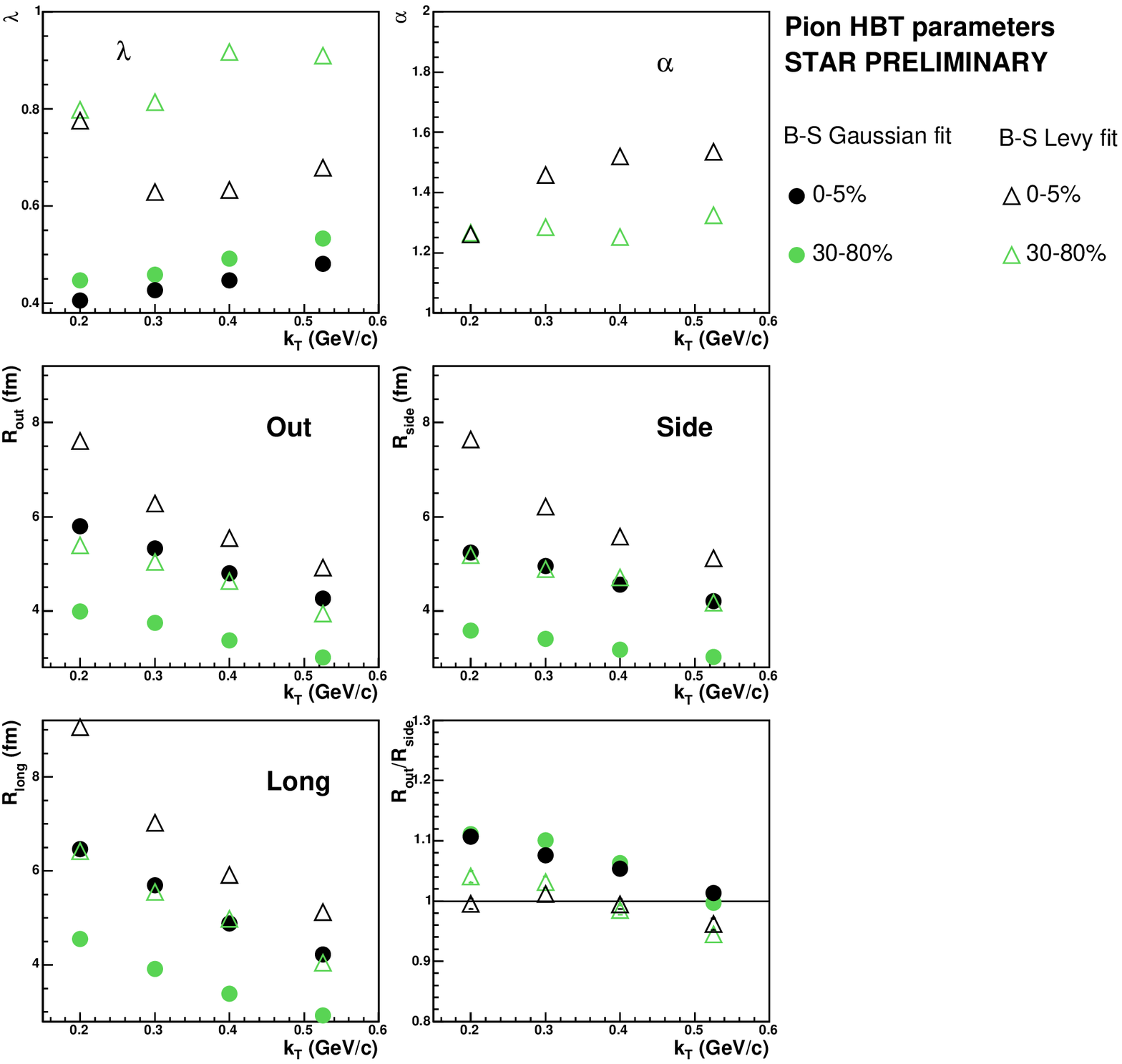}
  \caption{
    Results on interferometric parameters 
    as functions of $k_{\rm T}$ and centrality obtained using 
    Gaussian parametrized fit~\eqref{eq:Gaussian} 
    compared to the parameters of 
    Lévy stable source parametrized fit~\eqref{eq:Levy} of the 
    measured correlation function.}
  \label{BS-Levy-Parameters}
\end{figure}

In Figure~\ref{BS-Levy-Parameters} the 
interferometric parameters obtained from the 
Gaussian parametrization~\eqref{eq:Gaussian}
are compared to the parameters obtained using 
Lévy parametrization~\eqref{eq:Levy} of the 
measured three-dimensional correlation function.
Interferometric parameters are shown 
as functions of $k_{\rm T}$ for 
the most central and peripheral collisions.
The Lévy fit returns significantly
larger values of the fit parameters 
$R_{\rm o}$, $R_{\rm s}$, $R_{\rm l}$ and $\lambda$.
However, these parameters can not 
be directly compared to the Gaussian ones
representing the interferometric 
radii, because the non-Gaussian parameters
do not satisfy the definition of being the
widths of the source emission function.
The large values of the non-Gaussian fit parameters are 
strongly anti-correlated with
rather low value of the
Lévy index of stability $\alpha$,
which stays between $1.2 < \alpha < 1.5$.

It can be seen that the Lévy stable 
source distribution parametrization~\eqref{eq:Levy}
does not bring an advantage in describing the detail 
shape of the measured correlation function,
nor in the number of the fitting parameters.
Therefore use of the standard 
Gaussian parametrization~\eqref{eq:Gaussian}
is sufficient and preferred.

\section{Summary}
The preliminary results on identical pion interferometry
using high statistics sample of
Au+Au collisions at $\sqrt{s_{{\rm NN}}}=200~{\rm GeV}$
from the STAR experiment at RHIC have been presented.

The results of the Gaussian fit to the measured
three-dimensional correlation function are 
consistent within errors with the previously analyzed STAR data.

It has been shown that in the low relative momentum region the
Lévy stable source parametrization
does not fit the experimental correlation function
significantly better when
compared to standard Gaussian parametrization.

Edgeworth expansion provides the detailed 
fit to the measured correlation function~\cite{Adams:2004yc}, 
but the interpretation of the higher order fit
parameters is not clear.

It seems that to represent the experimental correlation 
function in Au+Au collisions at RHIC, the Gaussian 
parametrization is sufficient.


\begin{theacknowledgments}
This work is supported by the IRP~AV0Z10480505
and the Grant Agency of the Czech Republic
under contract 202/04/0793.
\end{theacknowledgments}

\bibliographystyle{aipprocl} 

\begin{thebibliography}{9}

\bibitem{Csorgo:2003uv}
T.~Csörg\H o, S.~Hegyi and W.~A.~Zajc,
{\em Bose-Einstein correlations for Levy stable source distributions},
Eur.\ Phys.\ J.\ C {\bf 36}, 67 (2004)
[arXiv:nucl-th/0310042].

\bibitem{Wiedemann:1999qn}
U.~A.~Wiedemann and U.~W.~Heinz,
{\em Particle interferometry for relativistic heavy-ion collisions},
Phys.\ Rept.\  {\bf 319}, 145 (1999)
[arXiv:nucl-th/9901094].

\bibitem{Csorgo:2000pf}
T.~Csörg\H o and S.~Hegyi,
{\em Model independent shape analysis of correlations in 1, 2 or 3 dimensions},
Phys.\ Lett.\ B {\bf 489}, 15 (2000).

\bibitem{Adams:2004yc}
J.~Adams {\it et al.}  [STAR Collaboration],
{\em Pion interferometry in Au + Au collisions at s(NN)**(1/2) = 200-GeV},
Phys.\ Rev.\ C {\bf 71}, 044906 (2005)
[arXiv:nucl-ex/0411036].

\bibitem{Podgoretsky:1982xu}
M.~I.~Podgoretsky,
{\em On The Comparison Of Identical Pion Correlations In 
Different Reference Frames},
Sov.\ J.\ Nucl.\ Phys.\ {\bf 37}, 272 (1983)
[Yad.\ Fiz.\  {\bf 37}, 455 (1983)].

\bibitem{Pratt:1986ev}
S.~Pratt,
{\em Coherence And Coulomb Effects On Pion Interferometry},
Phys.\ Rev.\ D {\bf 33}, 72 (1986).

\bibitem{Bertsch:1989vn}
G.~F.~Bertsch,
{\em Pion Interferometry As A Probe Of The Plasma},
Nucl.\ Phys.\ A {\bf 498}, 173C (1989).

\bibitem{Bowler:1991vx}
M.~G.~Bowler,
{\em Coulomb corrections to Bose-Einstein correlations have been greatly
exaggerated},
Phys.\ Lett.\ B {\bf 270}, 69 (1991).

\bibitem{Sinyukov:1998fc}
Y.~Sinyukov {\em et al.},
{\em Coulomb corrections for interferometry analysis of expanding hadron
systems},
Phys.\ Lett.\ B {\bf 432}, 248 (1998).
 
\end{thebibliography}

\end{document}